\documentclass[aps,prl,reprint,superscriptaddress,amsmath,amssymb,lengthcheck,longbibliography,showpacs]{revtex4-1}
\usepackage{graphicx}
\def\bra#1{\mathinner{\langle{#1}|}}
\def\ket#1{\mathinner{|{#1}\rangle}}
\def\braket#1{\mathinner{\langle{#1}\rangle}}
\begin{document}

\title{Object identification using correlated orbital angular momentum
  states}

\author{N\'estor Uribe-Patarroyo}
\email{uribepnr@bu.edu}
\affiliation{Dept. of Electrical and Computer Engineering, Boston
  University, 8 Saint Mary’s St., Boston, MA 02215, USA}
\author{Andrew Fraine}
\affiliation{Dept. of Electrical and Computer Engineering, Boston
  University, 8 Saint Mary’s St., Boston, MA 02215, USA}
\author{David S. Simon}
\affiliation{Dept. of Electrical and Computer Engineering, Boston
  University, 8 Saint Mary’s St., Boston, MA 02215, USA}
\affiliation{Dept. of Physics and Astronomy, Stonehill
  College, 320 Washington St., Easton, MA 02357, USA}
\author{Olga Minaeva}
\affiliation{Dept. of Biomedical Engineering, Boston University, 44
  Cummington St., Boston, MA 02215, USA}
\author{Alexander V. Sergienko}
\affiliation{Dept. of Electrical and Computer Engineering, Boston
  University, 8 Saint Mary’s St., Boston, MA 02215, USA}
\affiliation{Dept. of Physics, Boston University, 590
  Commonwealth Avenue, Boston, MA 02215, USA}

\begin{abstract}
Using spontaneous parametric down conversion as a source of entangled
photon pairs, correlations are measured between the orbital angular momentum (OAM) in a target beam (which contains an unknown object) and that in an empty reference beam. Unlike previous studies, the effects of the object on {\it off-diagonal} elements of the OAM correlation matrix are examined. Due to the presence of the object, terms appear in which the signal and idler OAM do not add up to that of the pump.
Using these off-diagonal correlations, the potential for high-efficiency object identification by means of
correlated OAM states is experimentally
demonstrated for the first time. The higher-dimensional OAM Hilbert space enhances the information capacity of this approach, while the presence of the off-diagonal correlations allows for recognition of
specific spatial signatures present in the object. In particular, this allows the detection of
discrete rotational symmetries and the efficient evaluation of multiple azimuthal
Fourier coefficients using fewer resources than in conventional pixel-by-pixel imaging. This
represents a demonstration of sparse sensing using
OAM states, as well as being the first correlated OAM experiment to measure properties of a real, stand-alone object, a necessary first step toward correlated OAM-based remote sensing.

\end{abstract}


\maketitle

\subsection{Introduction}

Correlated optical sensing uses various types of correlations between pairs of
photons or pairs of classical light beams to form an image or to detect specific object
features. This includes techniques such as ghost imaging
\cite{ghost_imaging1, ghost_imaging2, ghost_imaging3, ghost_imaging4} and compressive ghost
imaging \cite{CGI, CSCGI}. In ghost imaging, correlated light is sent
through two different paths, one of which contains the target object
and a bucket detector, and the other has a spatially-resolving
detector but no object. Correlation between the outputs of each detector allows
reconstruction of an image or sensing of particular features
of the object \cite{oam_filter_ghost_imaging}. In compressive sensing \cite{compressive_sensing1},
the illumination of the target object is modulated by a number of known
(usually random) spatially-varying transmission masks and the output
is collected by a bucket detector. An estimation of the object or its
features can be found by correlating the intensity of the detected
light with the mask profile. Compressive ghost imaging uses the
techniques of compressive sensing in a ghost imaging setup to attain
more efficient imaging, such as the recently proposed multiple-input
ghost imaging \cite{MIGI}, in which the bucket detector is replaced by
a sparse-pixelated detector. The importance of compressive sensing
stems from the fact that successful sensing of the object requires a
number of measurements much smaller than the image size (measured in number of
pixels), which has
important implications for more efficient remote-sensing techniques.

The use of orbital angular momentum (OAM) states in classical and
quantum imaging techniques has been shown to provide additional
effects that enhance the sensitivity to particular features of an
object.  For example, it has been shown that the use of OAM modes in
phase imaging configurations increases edge contrast by using a spiral
phase distribution as a filter \cite{oam_filter_ghost_imaging}.  In addition, digital
spiral imaging \cite{spiral_imaging1, spiral_imaging2} has been
proposed as a technique in which the OAM basis is used to illuminate
the object and to analyze the transmitted or reflected light.  The
two-dimensional spatial structure of the mode along with the high dimension of the
OAM basis set leads to the probing of two-dimensional objects without
obtaining a pixel by pixel reconstruction, analogous to the
approach of compressive sensing.

This high dimensionality combined with the use of
correlated two-photon states gives rise to an improved version of
digital spiral imaging which exploits the full two-dimensional OAM
joint spectrum. This new method \cite{CSI}, called correlated spiral imaging
(CSI), is potentially suitable for OAM-based remote sensing.
The two-photon joint OAM spectrum provides object information concerning spatial symmetries, yielding more efficient object
recognition than conventional imaging and sensing techniques.  CSI can
be viewed as similar to compressive sensing, in the sense that
sampling is performed in randomly-chosen elements of a high-dimensional OAM basis, so that if the
object has a compact representation in this basis, a small number of
measurements will suffice to identify it.

This letter presents, to the best of our knowledge, the first
experimental demonstration of a sparse sensing technique based on
OAM states. It uses a simplified CSI implementation that expands
states generated by spontaneous parametric
down conversion (SPDC) \cite{downconversion_oam1, downconversion_oam2,downconversion_oam_exp1}
in terms of the OAM basis in order to identify objects without attempting
image reconstruction.  Significant information from the object
is available by measuring the \emph{joint} OAM spectrum of two-photon
states, giving rise to novel features outside the main
OAM-conserving diagonal \cite{zeilinger}, due to the interaction with the
object.  In order to use this information for object identification,
the relationship of the joint OAM spectrum to the
azimuthal Fourier
coefficients is derived. In the field of remote sensing this capability enables
efficient object identification without the requirement of a full
two-dimensional image.  The present demonstration also represents the
first correlated OAM-based sensing experiment representative of a
practical remote-sensing setup, in which physical objects completely
detached from any optical component are used as targets for
identification. All previous experiments on correlated sensing using
OAM states (\cite{oam_filter_ghost_imaging} for example) have used
simulated objects drawn on a spatial light modulator (SLM). High sensitivity to the spatial
properties of the targets is obtained, due to the high amount of
information per photon attainable by the use of OAM states.

\subsection{Theoretical basis}

Correlated OAM states for the implementation of CSI are created
through the process of spontaneous parametric down conversion (SPDC).
The two-photon state produced by SPDC can be expressed as an expansion
in Laguerre-Gauss (LG) modes $\ket{l,p}$ \cite{downconversion_oam2}
defined as
\begin{equation}
  \braket{r,\phi|l,p} = k_p^{|l|} r^{|l|}
  e^{-r^2/w_0^2} \textrm{L}_p^{|l|}\left(2r^2/w_0^2\right) e^{i\phi
    l},
\end{equation}
where $k_p^{|l|}$ is a normalization constant, $L_n^\alpha(x)$ is the
generalized Laguerre polymonial of order $n$ and $w_0$ is the beam
waist as defined in this plane. The azimuthal phase term in the LG
expression gives rise to modes with quantized OAM of order $l$.
Discrimination between individual radial modes $p \neq 0$ is not
attainable by current experimental techniques
\cite{downconversion_oam3}, therefore only the case $p=0$ will be
considered for simplicity and the $p$ index will be dropped
hereafter. The following derivation is easily generalized for nonzero $p$.
In the case of detecting LG modes by the use of holograms
and mono-mode fibers \cite{zeilinger}, the cross-talk given by modes
with $p \neq 0$ should be taken into account
\cite{downconversion_oam2}. In the case of a detection scheme with
sensitivity to all $p$ modes \cite{downconversion_oam3}, projections
should consider both the $l$ and $p$ indices.

The total OAM content is known to be conserved in a collinear SPDC
process \cite{zeilinger}, and in the case of a Gaussian pump (OAM
content $l_p=0$) the two-photon state can be written as
\begin{equation}
  \ket{\Psi} = \sum_{l=-\infty}^\infty C_l \ket{-l}\ket{l},
\end{equation}
with $C_l$ given by the phase-matching condition and the relation
between the pump, signal and idler beam waists \cite{CSI,
  downconversion_oam2}. $C_l$ can be viewed as the \emph{natural} OAM
spectrum coefficients from SPDC.

\begin{figure}
  \includegraphics[width=\columnwidth]{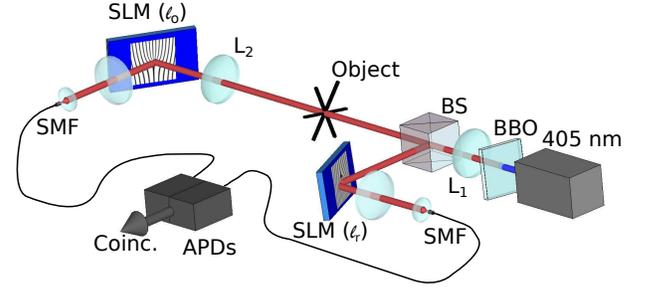}
  \caption{\label{fig:OptSetup} (Color online) Setup for the sensing
    of the object via orbital angular momentum of correlated photons.}
\end{figure}

Consider the SPDC setup of Fig.~\ref{fig:OptSetup}, where a correlated
OAM state is generated after the collinear down-converted beam enters
a non-polarizing beam splitter (BS). After splitting, one branch
corresponds to the reference arm and the other branch has the object
in its path. The stand-alone object is detached from any optical
component. The correlated state analyzer is composed of two identical
detection systems used to make the \emph{joint} projection on the LG
basis.  The detectors are connected to a coincidence circuit that
analyzes light with OAM number $l_r$ in the reference arm and $l_o$ in
the object arm. The coincidence rate is proportional to the
\emph{joint spectrum} $P(l_r,l_o)$ of the down-converted beam after
interaction with the object. This setup assumes the object is a
transmissive target; it is sufficient to reposition the detection
unit containing lens L2 in order to instead collect light in reflective mode from the
target.

The object is described by a spatially varying amplitude and/or
phase function $A(r, \phi)$. Interaction of an LG state with the
object can be expressed as an operator in the LG basis by the matrix
elements of the operator $A$
\begin{equation}
  \braket{k|A|l} = A_{kl}.
\end{equation}
One can calculate $\braket{k|A|l}$ (with $\rho = \sqrt{2} r / w_0$)
using
\begin{multline} \label{eq:Aproj}
\braket{k|A|l} = \sqrt{\frac{2}{\pi |k|!}}
  \sqrt{\frac{2}{\pi |l|!}} \times\\
  \frac{1}{2} \int_0^\infty d\rho
  \int_0^{2\pi} d\phi \rho^{|k|+|l|+1} e^{-\rho^2} e^{-i\phi(k-l)}
  A(\rho, \phi).
\end{multline}
By performing the integration in the radial direction the function
$R_{kl}(\phi)$ is obtained, which depends only on the azimuthal
variable
\begin{equation} \label{eq:int1}
  R_{kl}(\phi) = \frac{2}{\sqrt{|k|!|l|!}}
  \int_0^\infty d\rho \rho^{|k|+|l|+1} e^{-\rho^2} A(\rho, \phi).
\end{equation}
This integration reveals that $R_{kl}(\phi)$ corresponds to the
azimuthal features of $A(\rho, \phi)$ at a weighted average of object radius.
The weight function is the squared absolute value of an LG
mode $|l|+|k|$, which is an annular function with its maximum at a
radius $w_0\sqrt{(|l|+|k|)/2}$ \cite{oam_poynting}.

The second integration has the form (with $m=k-l$)
\begin{equation}\label{eq:int2}
  \braket{k|A|l} = \frac{1}{2 \pi} \int_0^{2\pi} d\phi
  e^{-im\phi} R_{kl}(\phi).
\end{equation}
Therefore, $A_{kl}$ is the $m$-th coefficient of the Fourier series of
$R_{kl}(\phi)$, the region selected by the ``ring'' in the first
integral. In the setup of Fig.~\ref{fig:OptSetup}, after interaction
with the object the two-photon state transforms into (dropping the
limits of the summations)
\begin{equation}
  \ket{\Psi'} = \sum_{l} C_{l} \ket{-l}\Big[A\ket{l}\Big] =
  \sum_{l} C_l \ket{-l}\left[\sum_{k'}A_{k'l}\ket{k'}\right].
\end{equation}
It follows that the joint spectrum is given by
\begin{equation}\label{eq:jointSpec}
  P(l_r,l_o) = \bra{l_r}\bra{l_o}|\Psi'\rangle = C_{l_r} A_{l_r,l_o},
\end{equation}
which contains information about the natural SPDC OAM spectrum
coefficients $C_l$ as well as information about the object
$A_{kl}$. The OAM spectral signature of the object can be isolated by
performing $A_{kl} = P(l,k) / C_l$ for $C_l \neq 0$.


States with $l_r + l_o \neq l_p$
can have non-zero probability. When there was no object present, these states were forbidden
by OAM conservation \cite{zeilinger, downconversion_oam1,
downconversion_oam2, downconversion_oam3, downconversion_oam_exp1}; they represent the interaction of the correlated OAM state from SPDC
with the object. These new non-conservation elements have a
contribution to its total OAM content from the object $m=l_o + l_r -
l_p$, and carry direct information of the $m$-fold rotational
symmetries of the object at a $|l_r| + |l_o|$ radius. For instance,
considering a Gaussian pump ($l_p=0$), the measured joint spectrum
with a $m$-fold rotationally symmetric target will show strong
components with total OAM content $\pm m$, and higher harmonics with
lower amplitudes at $nm$, $|n|>1$ being $n$ an integer.

\subsection{Experiment}

The experimental setup schematic is shown in
Fig.~\ref{fig:OptSetup}. A 1.5~mm-thick BBO crystal is pumped with a
30~mW diode laser at 405~nm.  To measure the OAM spectrum, an SLM is
used to display computer generated holograms.  The holograms change
the winding number $l$ of LG light, while two lenses and a mono-mode
fiber selectively couple modes with $l=0$ \cite{downconversion_oam2}.
One condition for obtaining $A_{kl}$ without any post-processing is
that the projections of Eqs. \eqref{eq:Aproj} and \eqref{eq:jointSpec}
are made in equivalent planes, i.e. planes that are imaged onto one another.
In the current setup, which is a
representative demonstration of remote sensing, lens L2 images an
object at a fixed distance onto the plane of the SLM.  In a general
remote sensing application, L2 will be placed appropriately so that all
targets will be at distances greater than the hyperfocal length of
L2. Therefore, all objects will be imaged in focus onto the SLM plane.

\begin{figure}[ht]
\begin{minipage}[c]{0.38\columnwidth}
  \includegraphics[width=0.9\columnwidth]{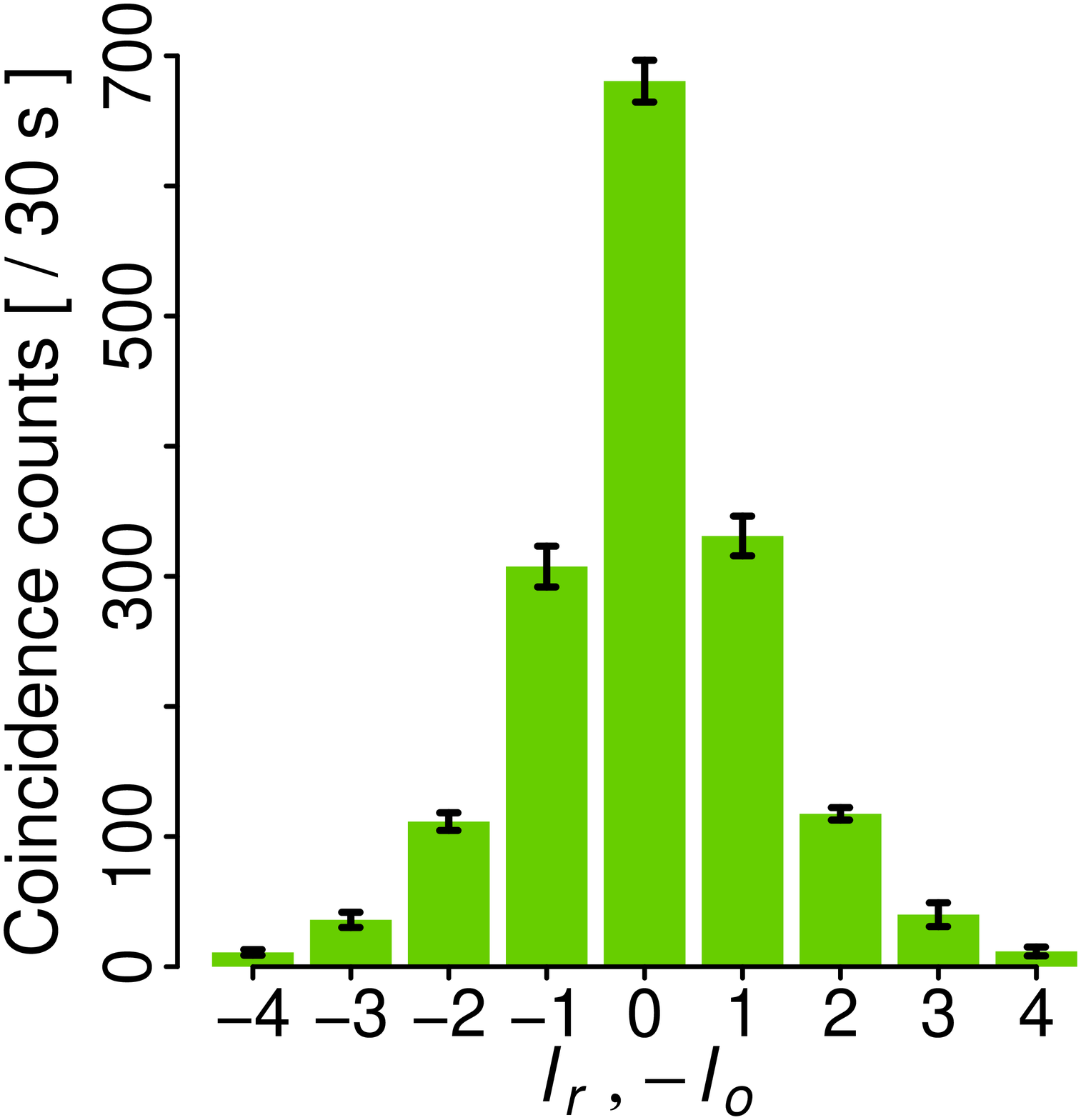}\\(a)
  \includegraphics[width=0.9\columnwidth]{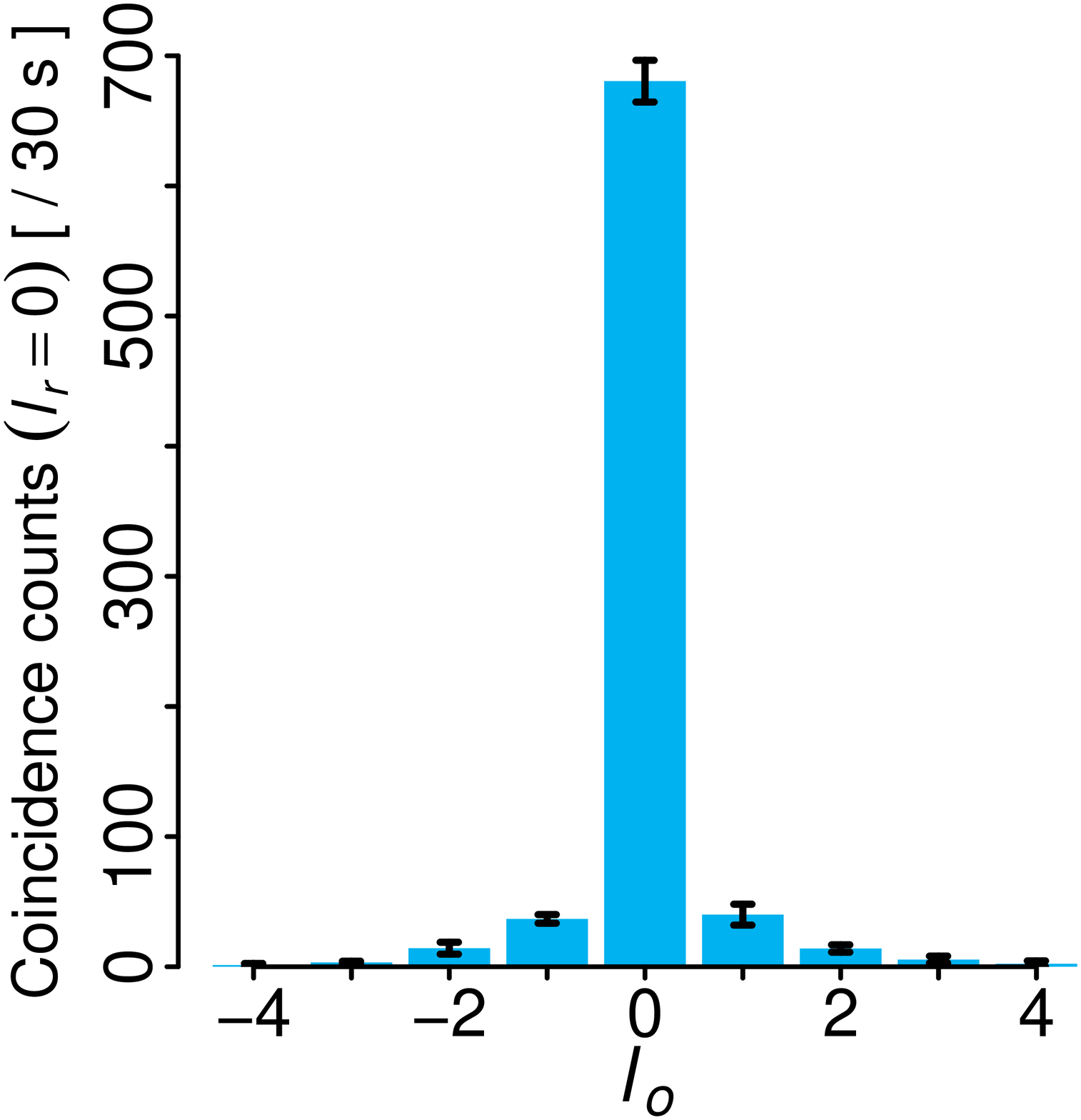}\\(c)
\end{minipage}
\begin{minipage}[c]{0.61\columnwidth}
  \includegraphics[width=\columnwidth]{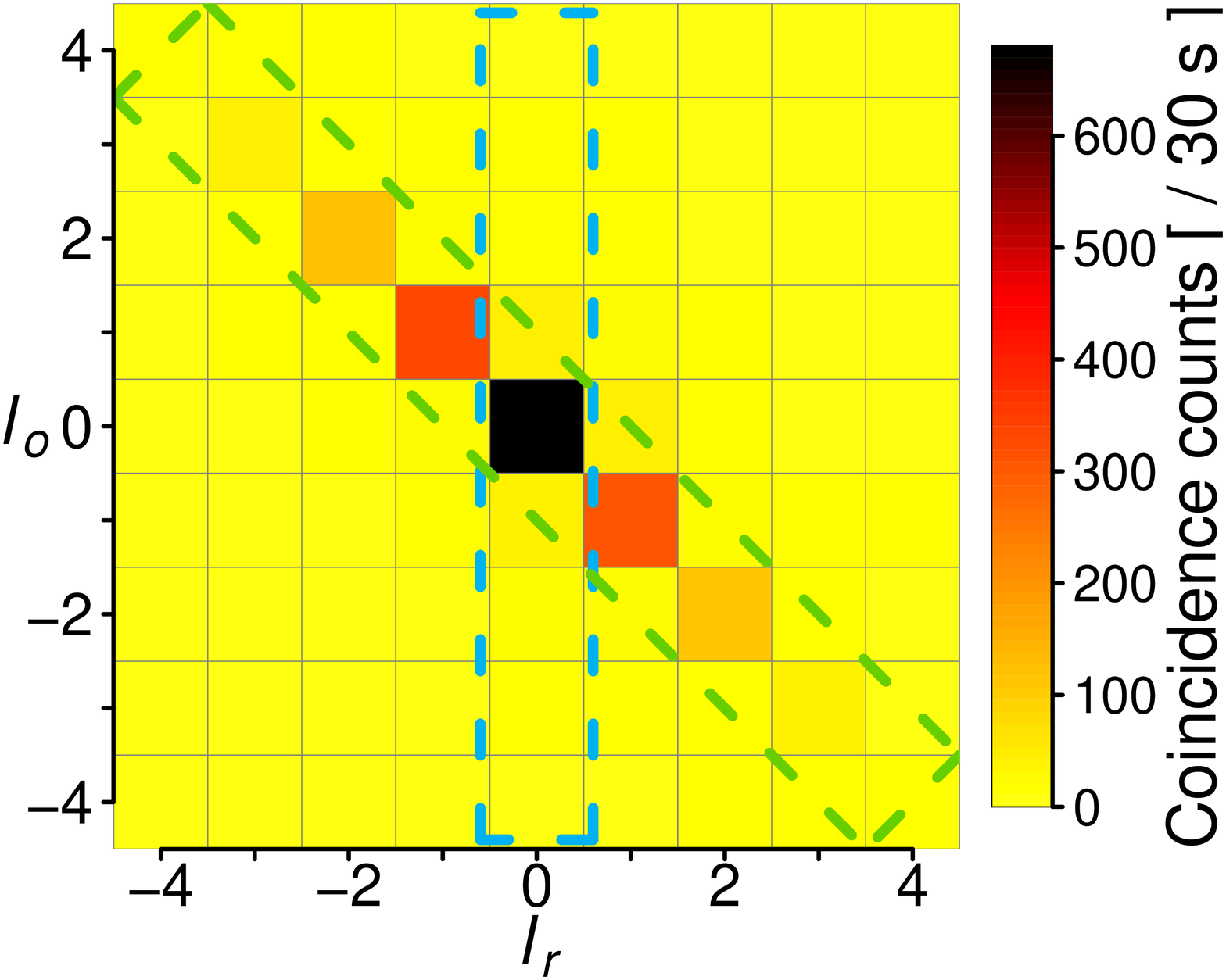}\\(b)
\end{minipage}
\caption{\label{fig:NoObjJS} (Color online) (a) Histogram of the main
  diagonal and (b) complete two-dimensional OAM joint spectrum for our
  configuration with no object. (c) Corresponding $P(l_o | l_r = 0)$
  cross section. The blue (vertical) box in the joint spectrum denotes
  the section in (a) and the green (diagonal) box denotes the section
  in (c).}
\end{figure}

Fig.~\ref{fig:NoObjJS} (a--b) illustrates the measurement of the
natural SPDC OAM joint spectrum $C_l$. The sign of $l_r$ is flipped to
compensate for the odd number of mirrors in order to reflect the OAM
conservation in SPDC.  In the absence of an object, only the main diagonal
terms which match the total OAM content of the pump appear, without any non-diagonal
terms observed \cite{downconversion_oam_exp1}.  Experimental data represent
the mean of four measurements, from which the standard deviation is
calculated and displayed as error bars. Fig.~\ref{fig:NoObjJS} (c)
shows the histogram of the diagonal and the cross section for $P(l_o |
l_r = 0)$.



To demonstrate the capability and high sensitivity of this new
technique for object recognition, the joint spectra using targets with
different rotational symmetries are analyzed.  As seen in
\eqref{eq:jointSpec}, an object can impart extra features in the joint
spectrum according to its azimuthal Fourier series at different radii.
In particular, an object with four-fold rotational symmetry will have
strong signatures corresponding to a total OAM content of $\pm 4$.  To
fit the scale of our setup, opaque strips 175~$\mu$m ($0.83w_0$)
thick, are placed in the object arm arranged with specific rotational
symmetries. These targets behave as transmission masks that block
light, and the integration time is adjusted in order to have a similar
amount of counts with respect to the case of no object.  Test objects
with well-defined dominant four- and six-fold symmetries shown in
Figs.~\ref{fig:2crossJS}~(a) and \ref{fig:3crossJS}~(a) considerably
modify the joint spectra [Figs.~\ref{fig:2crossJS}~(b) and
\ref{fig:3crossJS}~(b)] by adding extra diagonals with total OAM
content of $l = \pm 4$ and $l = \pm 6$, respectively.  The two objects
are clearly distinguished by the unique features of their respective
joint spectrum.

\subsection{Discussion}

\begin{figure}
  \begin{minipage}[c]{0.38\columnwidth}
    \includegraphics[width=0.7\columnwidth]{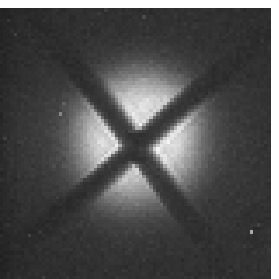}\\
    (a)\\
  \end{minipage}
  \begin{minipage}[c]{0.61\columnwidth}
    \includegraphics[width=0.85\columnwidth]{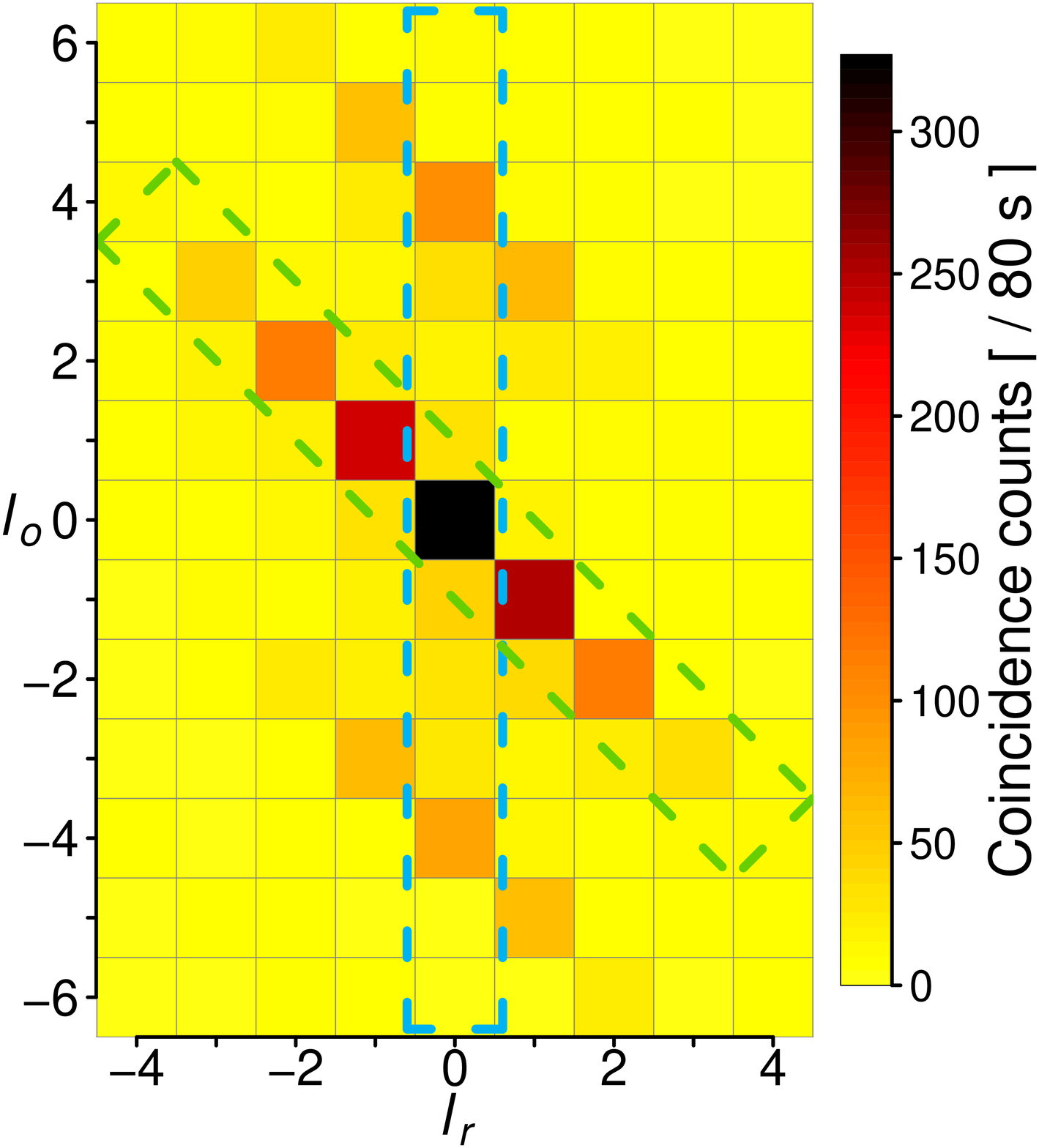}\\(b)
  \end{minipage}
  \begin{minipage}[c]{\columnwidth}
    \begin{minipage}[b]{0.44\columnwidth}
      \includegraphics[width=\columnwidth]{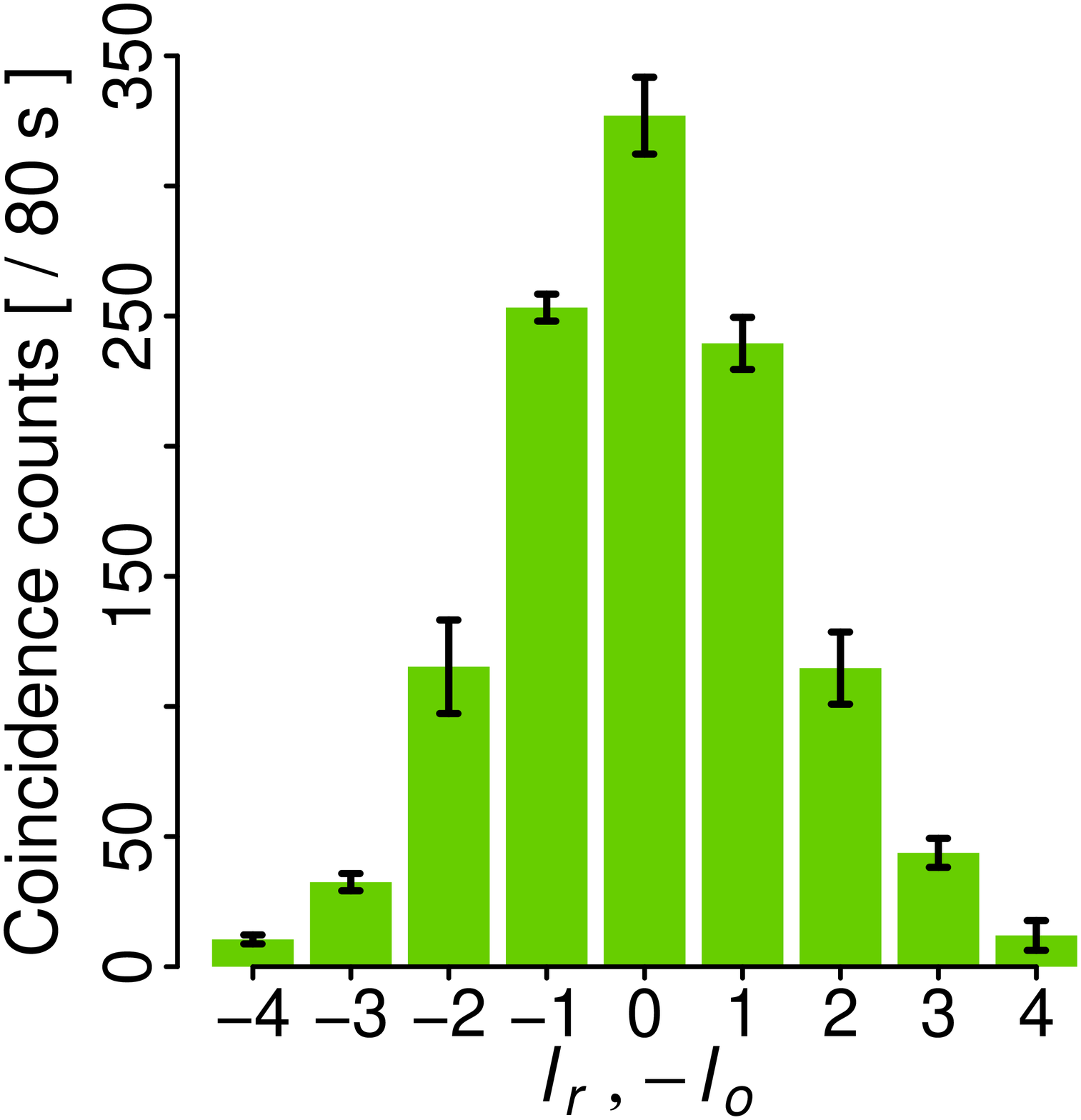}\\
      (c)
    \end{minipage}
    \begin{minipage}[b]{0.54\columnwidth}
      \includegraphics[width=\columnwidth]{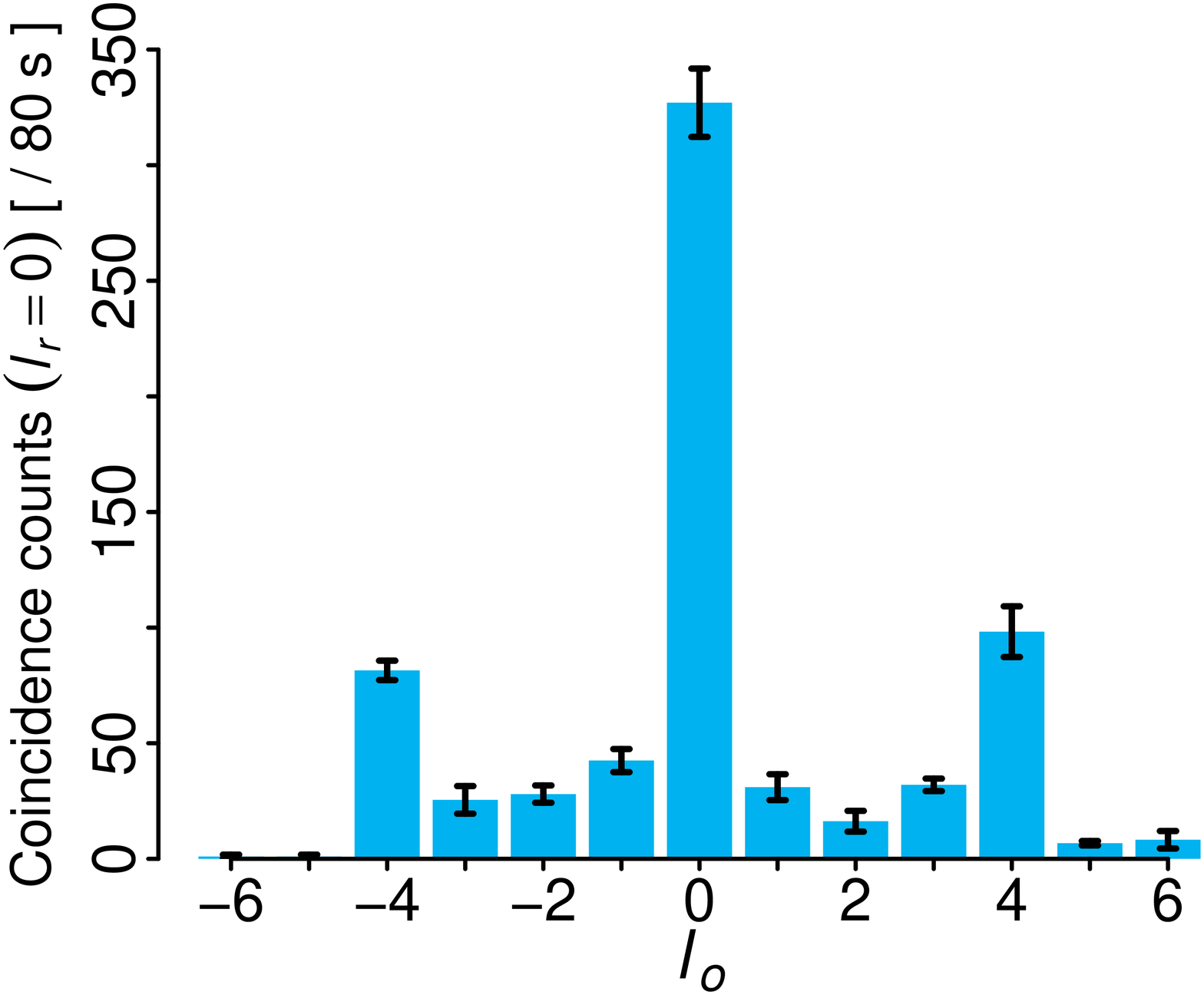}\\(d)
    \end{minipage}
  \end{minipage}
  \caption{\label{fig:2crossJS} (Color online) (a) Image of the cross
    used as target, (b) the experimental joint spectrum, (c) a
    histogram of the joint spectrum main diagonal and (d) $P(l_o | l_r
    = 0)$ cross section of the joint spectrum.}
\end{figure}

\begin{figure}
  \begin{minipage}[c]{0.38\columnwidth}
    \includegraphics[width=0.775\columnwidth]{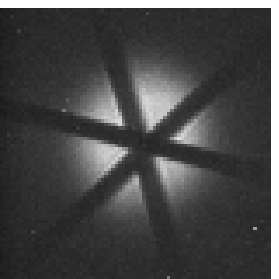}\\
    (a)\\

  \end{minipage}
  \begin{minipage}[c]{0.61\columnwidth}
    \includegraphics[width=0.85\columnwidth]{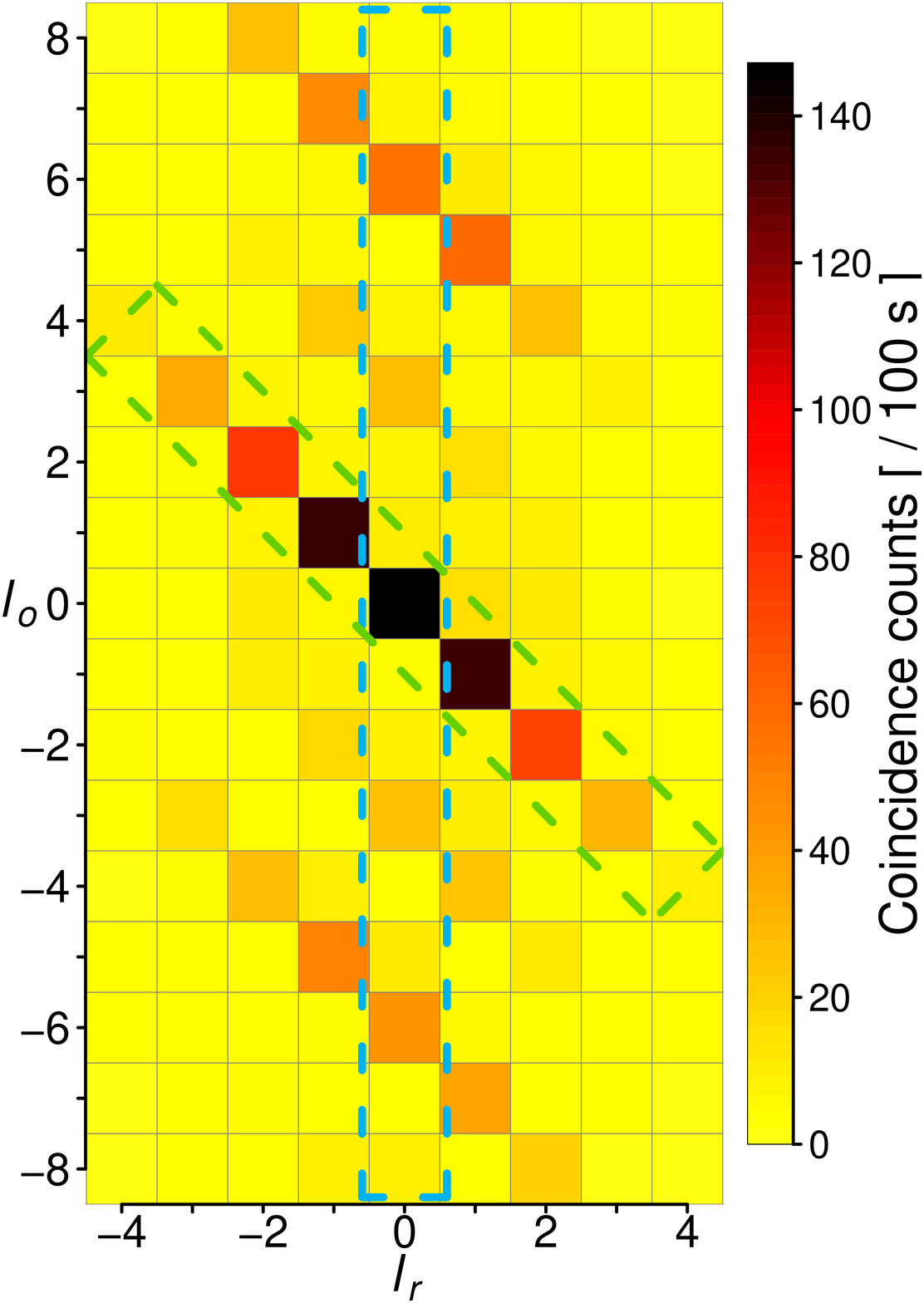}\\(b)
  \end{minipage}
  \begin{minipage}[c]{\columnwidth}
    \begin{minipage}[b]{0.44\columnwidth}
      \includegraphics[width=\columnwidth]{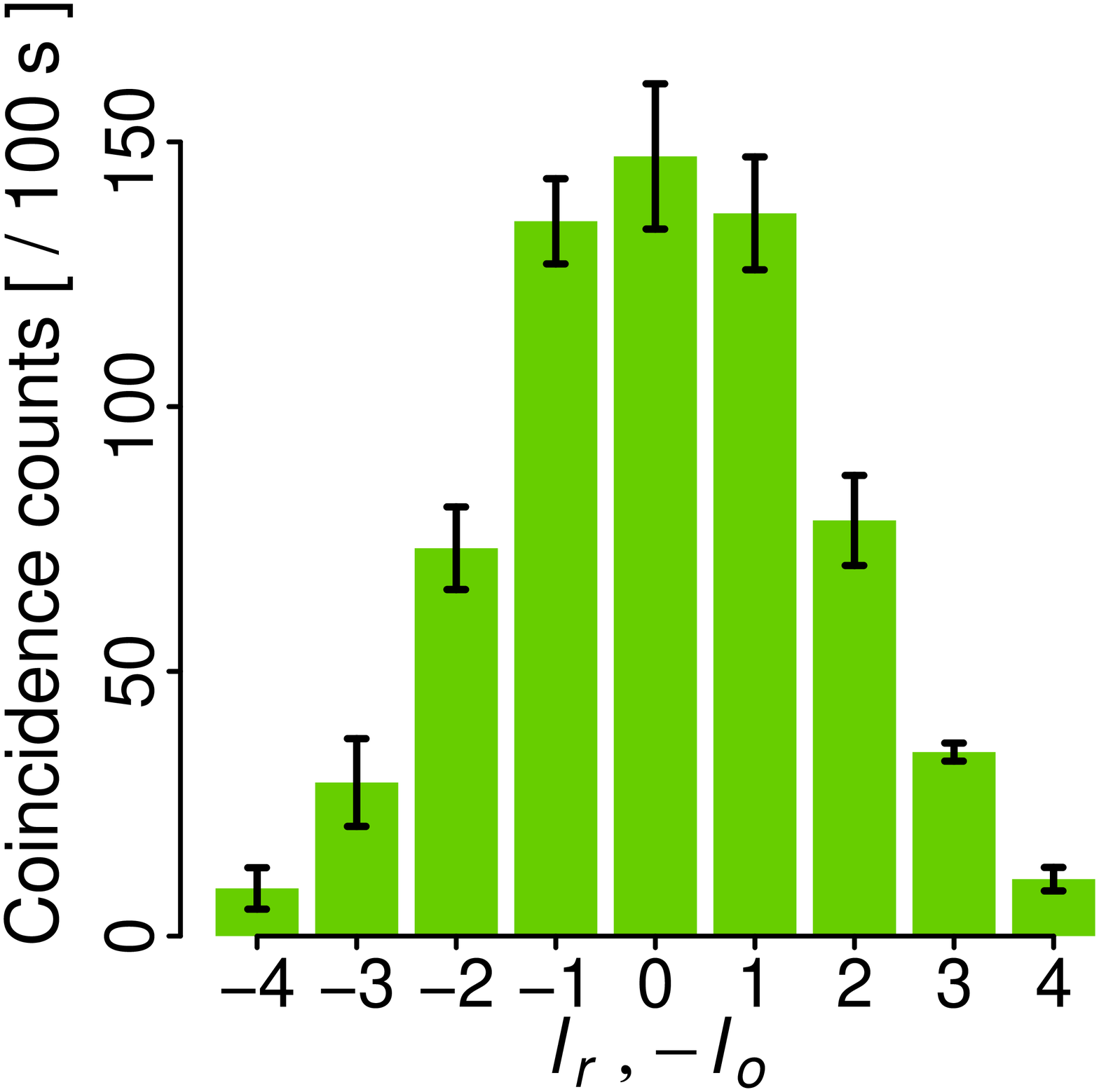}\\
    (c)
    \end{minipage}
    \begin{minipage}[b]{0.54\columnwidth}
        \includegraphics[width=\columnwidth]{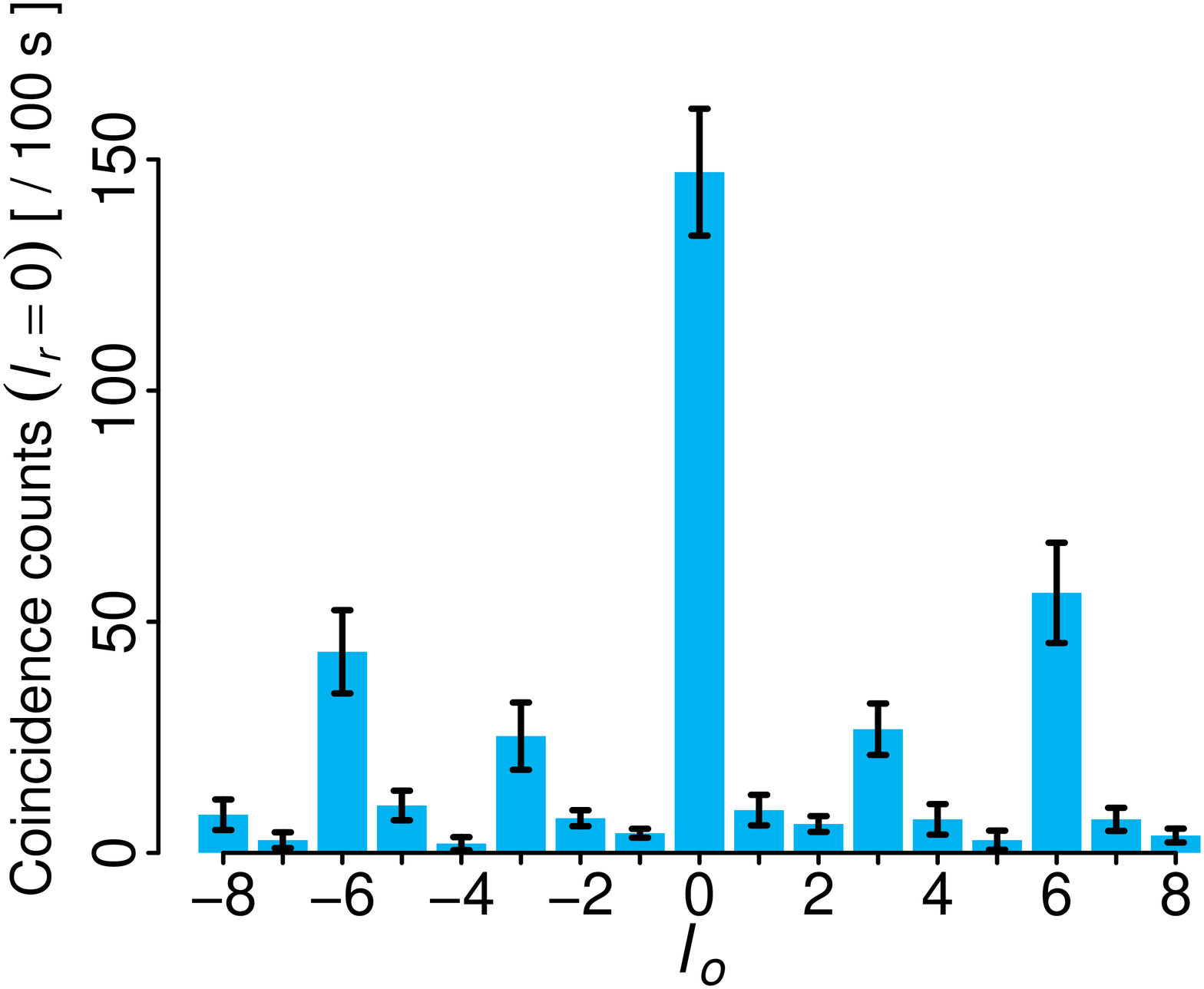}\\(d)
    \end{minipage}
  \end{minipage}
  \caption{\label{fig:3crossJS} (Color online) (a) Image of the
    three-arm cross used as target, (b) the experimental joint
    spectrum, (c) a histogram of the joint spectrum main diagonal and
    (d) $P(l_o | l_r = 0)$ cross section of the joint spectrum.}
\end{figure}

More insight about the structure of each object is acquired by
analyzing specific non-diagonal cross sections such as the two in
Figs.~\ref{fig:2crossJS}~(d) and \ref{fig:3crossJS}~(d).  Previous
works regarding the OAM joint spectrum of SPDC
\cite{downconversion_oam_exp1, downconversion_oam3, CSI,
  downconversion_oam1, downconversion_oam2} have considered only
the main diagonal elements corresponding to the conservation of OAM
with respect to the pump [Figs.~\ref{fig:2crossJS}~(c) and
\ref{fig:3crossJS}~(c)].  In the current case, there is an interaction
between the OAM from SPDC and the spatial features of the object
resulting in contributions which require analysis of the complete
two-dimensional joint spectrum.

The non-diagonal cross section $P(l_o | l_r = 0)$ shown in
Fig.~\ref{fig:2crossJS}~(d) indicates significant four-fold symmetry
as the dominant feature of the object.  This feature can be uniquely
attributed to the object because it is outside the diagonal that
corresponds to the conservation of OAM in SPDC.
Fig.~\ref{fig:3crossJS}~(d), showing the same non-diagonal cross
section for the six-fold symmetric object, exhibits richer features
associated with the higher complexity of the object.  Specifically,
apart from the dominant six-fold symmetric contribution, a three-fold
contribution is observed.  At the center of the object
[Fig.~\ref{fig:3crossJS}~(a)], the three stripes are displaced with
respect to the center. This forms a triangular shape in a small region
with three-fold rotational symmetry.  This small deviation from a
strict six-fold symmetry is readily observed in the non-diagonal cross
section by the appearance of significant contributions at $l_o = \pm
3$.

Although the cross section $P(l_o | l_r = 0)$ is a good indicator of
the presence of different symmetries, it is necessary to analyze the
full two-dimensional OAM joint spectrum to extract the relative
contribution of each symmetry component.  For example,
Fig.~\ref{fig:3crossJS}~(d) indicates a relative contribution of six-
and three-fold symmetries at a ratio of approximately 2:1.  However,
to judge the relative size of the areas within the object that exhibit
each symmetry, the total contribution of the entire diagonals should
be considered.  In this case, the joint spectrum in
Fig.~\ref{fig:3crossJS}~(b) clearly shows that the region with
three-fold symmetry is significantly smaller than the region with
six-fold symmetry.

As a final note, it should be pointed out that rotating the object about the azimuthal axis simply shifts the phase of the field by a constant, which will have no effect on the correlation rate. Therefore, if an object is rotated, the method will correctly continue to identify it as the same object.

In summary, a practical demonstration
of high-efficiency sensing using OAM states has been presented which allows
recognition of objects using a small number of measurements.
These results demonstrate that an object imprints its own
characteristic features onto the joint OAM joint spectrum of SPDC as
predicted by Eq.~\eqref{eq:jointSpec}.  New elements arise that do not
fulfill the OAM conservation condition required by SPDC alone, but which
are affected by the object as well.  The
capability of CSI for object recognition is demonstrated by these
results including high sensitivity to small symmetry components.  This
represents a realistic remote sensing application using physical
objects detached from any optical components.  Although the experiment was conducted with a transmissive object, a trivial modification
of the optical setup (simply changing the position of the lens, SLM, and object arm detector) makes it suitable for remote sensing of reflective targets at arbitrarily large distances.


\subsection{Acknowledgements}

This research was supported by the DARPA InPho program
through US Army Research Office award W911NF-10-1-0404.

\bibliography{OAM_correlated_sensing}

\end{document}